\documentclass[journal]{IEEEtran}
\IEEEoverridecommandlockouts
\usepackage{lineno}
\usepackage{hyperref}
\usepackage{cite}
\usepackage{amsmath,amssymb,amsfonts,cases} 
\usepackage{amsmath}
\usepackage{amsthm}

\DeclareMathOperator*{\argmin}{argmin}
\usepackage{graphicx}
\usepackage{textcomp}
\usepackage{xcolor}
\usepackage{graphicx}
\usepackage{float}
\usepackage{subfigure}
\usepackage{amsmath,mleftright}
\usepackage{amsfonts,amssymb}
\usepackage{mathrsfs}
\usepackage{mathtools}
\usepackage{algorithm}
\usepackage{algorithmic}
\usepackage{bm}
\usepackage{multirow}
\usepackage{array}
\usepackage{amssymb}
\usepackage{amsmath}
\usepackage{cite}
\usepackage{url}
\usepackage{xcolor}
\usepackage{cite,graphicx,amsmath,amssymb}
\usepackage{subfigure}
\usepackage{fancyhdr}
\usepackage{mdwmath}
\usepackage{mdwtab}
\usepackage{caption}
\usepackage{amsthm}
\usepackage{setspace}
\usepackage{bm}
\usepackage{mathtools}
\usepackage{dsfont}
\usepackage{bbm}
\usepackage{framed}
\newtheorem{remark}{Remark}
\newtheorem{theorem}{Theorem}

\newtheorem{lemma}{Lemma}

\newtheorem{corollary}{Corollary}

\makeatletter
\newcommand{\biggg}{\bBigg@{3}}
\newcommand{\Biggg}{\bBigg@{3.5}}
\makeatother
\makeatletter
\renewcommand{\maketag@@@}[1]{\hbox{\m@th\normalsize\normalfont#1}}%
\makeatother
\def\BibTeX{{\rm B\kern-.05em{\sc i\kern-.025em b}\kern-.08em
    T\kern-.1667em\lower.7ex\hbox{E}\kern-.125emX}}
    \expandafter\def\expandafter\normalsize\expandafter{%
    \normalsize%
    \setlength\abovedisplayskip{4pt}%
    \setlength\belowdisplayskip{4pt}%
    \setlength\abovedisplayshortskip{2pt}%
    \setlength\belowdisplayshortskip{2pt}%
}
\allowdisplaybreaks[4]

\begin{document}
\title{From Choirs to Solos: Reducing Multi-Cell Cooperation via Pinching Antennas}
\author{Zhenqiao Cheng, Chongjun~Ouyang, and Zhiguo~Ding,~\IEEEmembership{Fellow,~IEEE}\vspace{-10pt}}
\maketitle
\begin{abstract}
This article investigates the potential of the pinching-antenna system (PASS) to reduce multi-cell cooperation requirements. Flexible pinching-antenna (PA) placement can enhance desired channel gains and reduce the required transmit power, which consequently alleviates inter-cell interference and the need for cooperative transmission. A two-cell PASS is considered, where the cooperation cost is quantified by the number of additional base station-user data associations. The cooperation cost is minimized subject to user signal-to-interference-plus-noise ratio requirements and individual transmit-power constraints. Joint PA placement and transmission designs are developed using alternating optimization and second-order cone programming, while low-complexity zero-forcing-based alternatives are also proposed. Numerical results demonstrate that PASS substantially reduces cooperation requirements compared with the conventional fixed-antenna system. These findings highlight the potential of PASS to transform cooperative multi-cell ``\emph{choirs}'' into independent ``\emph{solos}''.
\end{abstract}

\section{Introduction}
\label{sec:introduction}

The pinching-antenna system (PASS) has recently attracted considerable attention due to its ability to enhance wireless channel gains \cite{liu2025pinching}. PASS employs dielectric waveguides to convey radio-frequency signals and pinching antennas (PAs) attached to the waveguides to radiate these signals into free space \cite{suzuki2022pinching}. By flexibly adjusting the PA positions along the waveguides, PASS provides a much larger spatial reconfiguration range than conventional reconfigurable antenna systems. In particular, a PA can be placed close to its intended user to establish a favorable line-of-sight (LoS) link and mitigate propagation loss and signal blockage \cite{suzuki2022pinching,liu2025pinching}.

These advantages have motivated extensive research on PASS, as summarized in recent surveys \cite{liu2025pinchingtutorial,lu2026survey,liu2025pinchingsurvey}. Most existing studies have focused on single-cell systems and investigated various physical-layer transmission designs. More recently, researchers have extended PASS to multi-cell networks. For example, stochastic-geometry-based studies have characterized network throughput \cite{zhu2025topological,sun2025stochastic}, while other works have investigated joint precoding, PA placement, and base station (BS)-user association \cite{chen2026pinching,li2026pinching,jiang2026pinching}. Despite these advances, the potential of PASS to reduce inter-cell cooperation requirements remains largely unexplored.

In conventional fixed-antenna systems, serving cell-edge users often requires high transmit power due to large propagation distances. The resulting inter-cell interference (ICI) further increases the power needed to satisfy users' quality-of-service (QoS) requirements in neighboring cells. A conventional solution is cooperative transmission, where multiple BSs jointly transmit user data through cooperative precoding. However, this approach incurs additional overhead associated with user-data sharing, cooperative signal processing, and inter-BS synchronization. PASS offers an alternative through flexible PA placement. By placing PAs close to their intended users, PASS shortens the wireless propagation distances and establishes stronger LoS links \cite{liu2025pinching}. The enhanced desired channel gains reduce the required transmit power \cite{ding2026toward} and consequently alleviate ICI \cite{ding2026characterizing}. Moreover, flexible PA placement can reshape the interference channels between neighboring cells. These advantages motivate an important question:
\emph{Can PASS reduce or even eliminate the need for cooperative transmission while satisfying users' QoS requirements?}

Motivated by this question, we investigate cooperation reduction in multi-cell PASS through joint PA placement and transmission design. We consider a representative two-cell system, where each cell employs one waveguide with one active PA to serve its associated user. We quantify the cooperation cost by the number of additional BS-user data associations and aim to minimize this cost subject to user signal-to-interference-plus-noise ratio (SINR) and individual BS power constraints. To this end, we investigate three transmission modes: independent transmission, one-sided cooperation, and full cooperation. For each mode, we jointly optimize PA placement and transmit power or cooperative precoding to minimize the required peak transmit power. We develop an alternating optimization (AO) framework with second-order cone programming (SOCP) for the cooperative designs and further propose low-complexity zero-forcing (ZF)-based alternatives. Numerical results demonstrate that PASS substantially reduces cooperation requirements compared with the conventional fixed-antenna system and enables independent per-cell transmission even under stringent QoS requirements. These findings highlight the potential of PASS to transform multi-cell cooperative \emph{``choirs''} into independent \emph{``solos''}.

\section{System Model}
\label{sec:system}

\begin{figure}[!t]
 \centering
\includegraphics[width=0.4\textwidth]{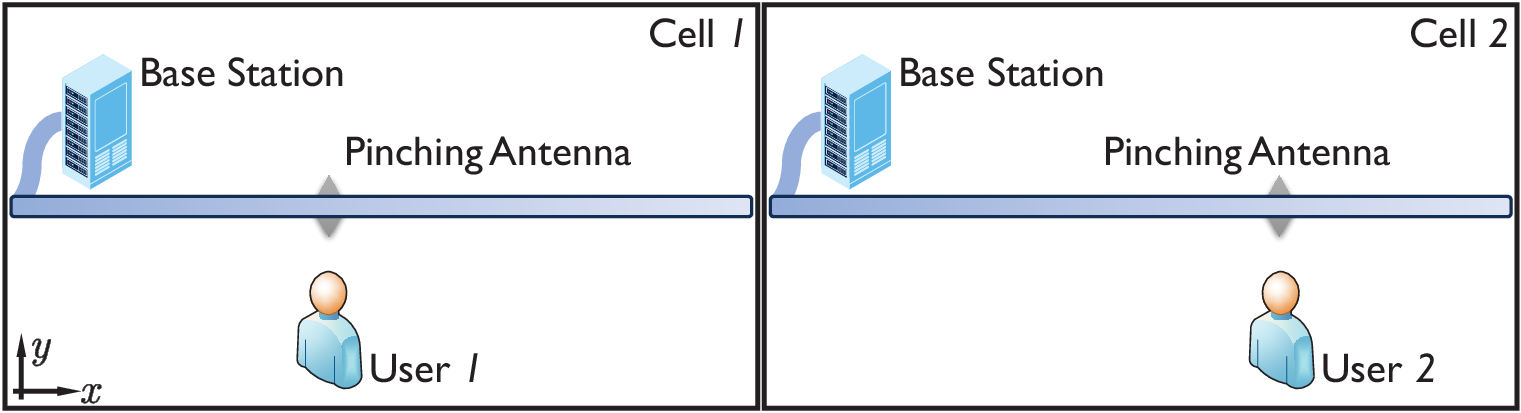}
\caption{Illustration of a two-cell PASS.}
\label{Figure: System_Model}
\vspace{-15pt}
\end{figure}

Consider the two-cell PASS illustrated in {\figurename} {\ref{Figure: System_Model}}. Each cell $k\in\{1,2\}$ has length $D_x$ along the $x$-axis and width $D_y$ along the $y$-axis. The center of cell $k$ is located at $(x_k^{\rm C},y_k^{\rm C})$, and its service region is defined as $
\mathcal{C}_k
\triangleq
[x_k^{\rm{C}}-\frac{D_x}{2},x_k^{\rm{C}}+\frac{D_x}{2}]
\times
[y_k^{\rm{C}}-\frac{D_y}{2},y_k^{\rm{C}}+\frac{D_y}{2}]$. Each BS is equipped with a dielectric waveguide deployed along the $x$-axis at height $H>0$ above the user plane. The waveguide spans the entire cell length, and one PA is activated on each waveguide. Let $t_k\in[-\frac{D_x}{2},\frac{D_x}{2}]\triangleq\mathcal{T}$ denote the PA position relative to the center of cell $k$. Its three-dimensional location is given by $\mathbf{t}_k(t_k)=[x_k^{\rm C}+t_k,y_k^{\rm C},H]^{\mathsf T}$. User $k$ is located at $\mathbf{u}_k=
[x_k^{\rm U},y_k^{\rm U},0]^{\mathsf T}$ in cell $k$ and is primarily associated with BS $k$. Both users are simultaneously served over the same time-frequency resource. A central controller has perfect knowledge of the user locations and position-dependent channel coefficients and jointly determines the PA positions and transmission configurations of both cells. The user locations and propagation environment remain unchanged during each scheduling period.

\subsection{Channel Model}
\label{subsec:channel}

PASS is particularly attractive for high-frequency communications \cite{suzuki2022pinching}, where LoS propagation typically dominates \cite{ouyang2024primer}. We thus consider LoS propagation between the PAs and users. The free-space channel coefficient from the PA in cell $k'$ to user $k$ is modeled as follows \cite{ouyang2024primer}:
\begin{align}
h_{k',k}^{\rm LoS}(t_{k'})
=
\frac{\sqrt{\eta}\,
{\rm e}^{-{\rm j}k_0
\|\mathbf{t}_{k'}(t_{k'})-\mathbf{u}_k\|}}
{\|\mathbf{t}_{k'}(t_{k'})-\mathbf{u}_k\|},
\nonumber
\end{align}
where $\eta\triangleq\frac{\lambda^2}{16\pi^2}$, $\lambda$ is the free-space wavelength, and $k_0\triangleq\frac{2\pi}{\lambda}$ is the free-space wavenumber. In addition to free-space propagation,
the signal propagates along the dielectric
waveguide from the feed point to the
activated PA.
Let $\mathbf{r}_{k'}^{\rm Feed}$ denote
the feed-point location of waveguide $k'$.
The corresponding in-waveguide channel
coefficient is modeled as follows:
\begin{align}
h_{k'}^{\rm In}(t_{k'})
=
{\rm e}^{-{\rm j}n_{\rm eff}k_0
\|\mathbf{t}_{k'}(t_{k'})
-\mathbf{r}_{k'}^{\rm Feed}\|},
\nonumber
\end{align}
where $n_{\rm eff}$ is the effective
refractive index of the dielectric waveguide. The end-to-end channel coefficient from
BS $k'$ to user $k$ is therefore given by
$h_{k',k}(t_{k'})
=h_{k'}^{\rm In}(t_{k'})
h_{k',k}^{\rm LoS}(t_{k'})$.
Accordingly, the channel power gain is
expressed as follows:
\begin{align*}
g_{k',k}(t_{k'})
\triangleq |h_{k',k}(t_{k'})|^2
=
\frac{\eta}
{q_{k',k}+(t_{k'}-\xi_{k',k})^2},
\end{align*}
where
$q_{k',k}\triangleq
(y_k^{\rm U}-y_{k'}^{\rm C})^2+H^2>0$
and
$\xi_{k',k}\triangleq
x_k^{\rm U}-x_{k'}^{\rm C}$.

\subsection{Cooperative Transmission}
\label{subsec:multicell_transmission}

\begin{figure}[!t]
 \centering
\includegraphics[width=0.4\textwidth]{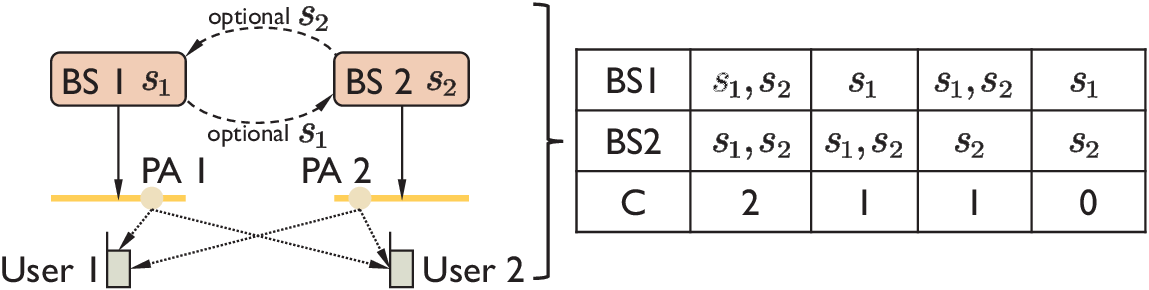}
\caption{Illustration of two-cell cooperation.}
\label{Figure: Cooperationl_BS}
\vspace{-15pt}
\end{figure}

Let $s_k$ denote the unit-power information symbol intended for user $k$. When cooperation is enabled, each BS may also transmit the neighboring user's data, as illustrated in {\figurename} {\ref{Figure: Cooperationl_BS}}. Let $w_{k',k}$ denote the digital precoding coefficient applied to $s_k$ at BS $k'$. The transmitted signal from BS $k'$ is given by $x_{k'}=\sum_{k=1}^{2}w_{k',k}s_k$. In particular, $w_{k',k}=0$ for $k'\neq k$ indicates that BS $k'$ does not transmit user $k$'s data. The received signal at user $k$ is expressed as follows:
\begin{align*}
y_k=\sum\nolimits_{k'=1}^{2}h_{k',k}(t_{k'})
\sum\nolimits_{j=1}^{2}w_{k',j}s_{j}+n_k,
\end{align*}
where $n_k\sim\mathcal{CN}(0,\sigma^2)$ denotes the
additive noise with power $\sigma^2$.
Coherent inter-BS phase synchronization is assumed
whenever joint transmission is activated.
The SINR
of user $k$ is given by
\begin{align*}
\Gamma_k=
\frac{
|\sum_{k'=1}^{2}
h_{k',k}(t_{k'})w_{k',k}|^2
}{
|\sum_{k'=1}^{2}
h_{k',k}(t_{k'})w_{k',j}|^2
+\sigma^2
},
\end{align*}
where $j\ne k$. In conventional fixed-antenna systems, serving cell-edge users often requires high transmit power, which can cause substantial ICI. Cooperative precoding can mitigate ICI but incurs additional overhead due to inter-BS data sharing, signal processing, and synchronization.
\vspace{-5pt}
\begin{remark}
PASS can substantially shorten the wireless propagation distances through flexible PA placement, which enhances the desired channel gains and reduces the required transmit power \cite{ding2026toward}. The reduced transmit power alleviates ICI \cite{ding2026characterizing}. Consequently, PASS may satisfy the same QoS requirements with less cooperative transmission than conventional fixed-antenna systems.
\end{remark}
\vspace{-5pt}
Define the cooperative precoding matrix as follows:
\begin{align*}
\mathbf W\triangleq
\begin{bmatrix}
w_{1,1}&w_{1,2}\\
w_{2,1}&w_{2,2}
\end{bmatrix}
\in\mathbb C^{2\times2}.
\end{align*}
Under independent transmission, each BS transmits only
its own user's data, and $\mathbf W$ is diagonal.
When $w_{k',k}\neq0$ for $k'\neq k$, user $k$'s data
must additionally be shared with BS $k'$.
We therefore define the cooperation cost as follows:
\begin{align*}
C(\mathbf W)\triangleq
\mathbbmss{1}\{|w_{1,2}|>0\}
+\mathbbmss{1}\{|w_{2,1}|>0\}
\in\{0,1,2\},
\end{align*}
where $\mathbbmss{1}\{\cdot\}$ denotes the indicator function. Moreover, $C=0$, $C=1$, and $C=2$ correspond to independent transmission, one-sided cooperation, and two-sided cooperation, respectively, as illustrated in {\figurename} {\ref{Figure: Cooperationl_BS}}. 
\vspace{-5pt}
\begin{remark}
The cooperation cost \(C\) quantifies the number of additional cross-cell user-data associations beyond the primary BS-user associations. For an equal scheduled payload of \(L\) bits per user, the associated additional user-data delivery requirement is \(LC\) bits per scheduling interval. In particular, \(C=0\) eliminates additional cross-cell user-stream delivery and the requirement for coherent joint transmission. The proposed metric therefore captures a concrete component of cooperation-related operational burden.
\end{remark}
\vspace{-5pt}

\subsection{Problem Formulation}
\label{subsec:rb_problem}

Our objective is to minimize the cooperation cost required
to satisfy both users' QoS requirements under a per-BS
transmit power budget $P_{\max}$.
Let $\mathbf t=[t_1,t_2]^{\mathsf T}$ and let $\gamma$
denote the common SINR target.
The optimization problem can be formulated as follows:
\begin{subequations}
\label{eq:main}
\begin{align}
\min_{\mathbf t,\mathbf W}\quad
& C(\mathbf W)
\label{eq:main_obj}\\
\mathrm{s.t.}\quad
&t_k\in\mathcal T,~\Gamma_k\geq\gamma,\quad k=1,2,
\label{eq:main_sinr}\\
&\sum\nolimits_{k'=1}^{2}|w_{k,k'}|^2
\leq P_{\max},\quad k=1,2.
\label{eq:main_power}
\end{align}
\end{subequations}
To solve problem~\eqref{eq:main}, we consider three
cooperation levels $c\in\{0,1,2\}$ and define the corresponding feasible precoding sets as follows:
\begin{align*}
\mathcal W_0
&\triangleq
\{\mathbf W:w_{1,2}=w_{2,1}=0\},\\
\mathcal W_1
&\triangleq
\{\mathbf W:w_{1,2}=0
\ \text{or}\ w_{2,1}=0\},\\
\mathcal W_2
&\triangleq\mathbb C^{2\times2}.
\end{align*}
For each cooperation level $c$, the minimum required
peak transmit power is defined as follows:
\begin{align*}
Q_c^\star
\triangleq
\min_{\mathbf t,\mathbf W}\quad
&\max_{k=1,2}
\sum\nolimits_{k'=1}^{2}|w_{k,k'}|^2\\
\mathrm{s.t.}\quad
&{\mathbf W}\in\mathcal W_c,~\Gamma_k\geq\gamma,~t_k\in\mathcal T,~ k=1,2.
\end{align*}
We set $Q_c^\star=\infty$ if the corresponding
problem is infeasible.
Since
$\mathcal W_0\subseteq\mathcal W_1\subseteq\mathcal W_2$,
the minimum required powers satisfy $Q_2^\star\leq Q_1^\star\leq Q_0^\star$. Consequently, the minimum cooperation cost is determined by
\begin{align*}
C^\star=
\min\{c\in\{0,1,2\}:Q_c^\star\leq P_{\max}\},
\end{align*}
provided that $Q_2^\star\leq P_{\max}$.
Otherwise, the QoS requirements cannot be satisfied
even under full cooperation.

\section{Joint PA Placement and Transmission Design}
\label{sec:solution}

\subsection{No Cooperation: $C=0$}
\label{subsec:no_cooperation}

We first consider independent transmission, where
$w_{1,2}=w_{2,1}=0$.
Let $p_k=|w_{k,k}|^2$ denote the transmit power of BS $k$.
For fixed PA positions $\mathbf t$, the SINR constraints
reduce to
\begin{align*}
\frac{p_1g_{1,1}(t_1)}{\gamma}
\geq p_2g_{2,1}(t_2)+\sigma^2,
\frac{p_2g_{2,2}(t_2)}{\gamma}
\geq p_1g_{1,2}(t_1)+\sigma^2.
\end{align*}
The componentwise minimum feasible power solution is
obtained when both inequalities hold with equality.
For notational simplicity, the arguments of the channel
power gains are omitted hereafter.
Define
$\Delta_0\triangleq
g_{1,1}g_{2,2}-\gamma^2g_{1,2}g_{2,1}$.
When $\Delta_0>0$, the minimum required
transmit powers are 
\begin{align*}
p_1^\star(\mathbf t)
&=
\frac{\gamma\sigma^2
(g_{2,2}+\gamma g_{2,1})}
{\Delta_0},~
p_2^\star(\mathbf t)
=
\frac{\gamma\sigma^2
(g_{1,1}+\gamma g_{1,2})}
{\Delta_0}.
\end{align*}
Otherwise, no finite feasible power solution exists.
The minimum required peak transmit power under
independent transmission is thus given by
$Q_0^\star=\min_{t_1,t_2\in\mathcal T}
\Phi_0(\mathbf t)$, where
\begin{equation}
\Phi_0(\mathbf t)
\triangleq
\begin{cases}
\max\{p_1^\star(\mathbf t),p_2^\star(\mathbf t)\},
&\Delta_0>0,\\
\infty,&\text{otherwise}.
\end{cases}
\nonumber
\end{equation}

To address the coupling between the two PA positions, we adopt an element-wise AO method. Specifically, each PA position is optimized while the other is fixed:
\begin{align*}
\min\nolimits_{t_k\in\mathcal T}
\Phi_0(\mathbf t),\quad k\in\{1,2\}.
\end{align*}
Since each subproblem involves only one variable over
a bounded interval, we employ a one-dimensional search.
Specifically, the feasible interval
$\mathcal T=[-\frac{D_x}{2},\frac{D_x}{2}]$
is discretized into a $Q$-point grid
$\mathcal Q=
\{-\frac{D_x}{2}
+\frac{(q-1)D_x}{Q-1}
\}_{q=1}^{Q}$.
The PA position is then updated as follows:
\begin{align*}
t_k
\leftarrow
\argmin\nolimits_{t\in\mathcal Q\cup\{t_k\}}
\Phi_0(\mathbf t|_{t_k=t}),
\end{align*}
where $\mathbf t|_{t_k=t}$ denotes the PA-position vector
with its $k$th entry replaced by $t$.
Candidate positions that do not yield finite feasible
power solutions are discarded.
The two PA positions are updated sequentially until
convergence.

\subsection{One-Sided Cooperation: $C=1$}
\label{subsec:one_sided}

Under one-sided cooperation, at most one cross-cell
data stream is allowed, i.e., either $w_{1,2}=0$
or $w_{2,1}=0$.
We optimize the two possible cooperation directions
separately and select the one with the smaller required
peak transmit power.
For either direction, the joint design is addressed
through AO between digital precoding and PA placement.

\subsubsection{Digital Precoding Optimization}

For fixed PA positions $\mathbf{t}$, the minimum peak-power
precoding problem is formulated as follows:
\begin{align}\label{eq:cooperative_socp}
\min_{\mathbf W,P}~
P\quad
\mathrm{s.t.}~
\Gamma_k\geq\gamma,~\sum\nolimits_{k'=1}^{2}|w_{k,k'}|^2
\leq P,~ k=1,2,
\end{align}
together with the corresponding zero-entry constraint
on $\mathbf W$.
By rotating the phase of each user-stream precoder,
the SINR constraints can be equivalently expressed as follows:
\begin{align}
\sqrt{|\mathbf h_k^{\mathsf T}\mathbf w_j|^2+\sigma^2}
\leq
\frac{
\Re\{\mathbf h_k^{\mathsf T}\mathbf w_k\}
}{\sqrt{\gamma}},~
\Im\{\mathbf h_k^{\mathsf T}\mathbf w_k\}=0,~j\neq k,
\nonumber
\end{align}
where
$\mathbf h_k=[h_{1,k}(t_1),h_{2,k}(t_2)]^{\mathsf T}$
and
$\mathbf w_k=[w_{1,k},w_{2,k}]^{\mathsf T}$.
Thus, the precoding problem can be reformulated as an SOCP and solved optimally
for fixed $\mathbf t$ using convex optimization solvers
such as CVX.

\subsubsection{PA Placement Optimization}

For the optimized precoding matrix $\mathbf W$, define $\mathbf v_k=\frac{\mathbf w_k}{\sqrt{\hat p_k}}=[v_{1,k},v_{2,k}]^{\mathsf T}$, where $\hat p_k=\|\mathbf w_k\|^2$ denotes the total power allocated to user stream $k$. Let $\mathbf V=[\mathbf v_1,\mathbf v_2]$ denote the normalized precoding directions. For fixed $\mathbf V$, define the effective channel gains $\hat g_{k',k}\triangleq|\sum_{j=1}^{2}v_{j,k'}h_{j,k}(t_j)|^2$. The SINR constraints then reduce to
\begin{align*}
\frac{\hat p_1\hat g_{1,1}}{\gamma}
\geq \hat p_2\hat g_{2,1}+\sigma^2,\quad
\frac{\hat p_2\hat g_{2,2}}{\gamma}
\geq \hat p_1\hat g_{1,2}+\sigma^2.
\end{align*}
Define
${\hat\Delta}_0\triangleq
\hat g_{1,1}\hat g_{2,2}
-\gamma^2\hat g_{1,2}\hat g_{2,1}$.
For $\hat\Delta_0>0$, the minimum required stream powers
are given by
\begin{align}
\label{eq:cooperative_power}
\hat p_1^\star
&=
\frac{\gamma\sigma^2
(\hat g_{2,2}+\gamma\hat g_{2,1})}
{{\hat\Delta}_0},~
\hat p_2^\star
=
\frac{\gamma\sigma^2
(\hat g_{1,1}+\gamma\hat g_{1,2})}
{{\hat\Delta}_0}.
\end{align}

Accordingly, for fixed precoding directions $\mathbf V$, the minimum required peak BS power is expressed as follows:
\begin{equation*}
\Phi(\mathbf t;\mathbf V)
=
\max_{k=1,2}
\sum\nolimits_{k'=1}^{2}
\hat p_{k'}^\star|v_{k,k'}|^2,
\end{equation*}
where $\Phi(\mathbf t;\mathbf V)=\infty$ if $\hat\Delta_0\leq0$. We then optimize the two PA positions sequentially to minimize $\Phi(\mathbf t;\mathbf V)$ using the element-wise AO and one-dimensional search described in Section~\ref{subsec:no_cooperation}. After convergence, the stream powers are updated by \eqref{eq:cooperative_power}, and the precoding matrix is reconstructed as $\mathbf W=\mathbf V
\operatorname{diag}(\sqrt{\hat p_1^\star},
\sqrt{\hat p_2^\star})$. The digital precoding and PA placement updates are repeated until convergence. The procedure is performed for both cooperation directions, and the solution with the smaller required peak power is selected.

\subsection{Full Cooperation: $C=2$}
\label{subsec:full_cooperation}

Under full cooperation, both BSs may transmit both
users' data streams, and the precoding matrix
$\mathbf W\in\mathbb C^{2\times2}$ is unrestricted.
The AO algorithm developed for one-sided cooperation
directly applies to this case, except that the zero-entry
constraint on $\mathbf W$ is removed from
\eqref{eq:cooperative_socp}.

\subsection{ZF-Based Low-Complexity Design}
\label{subsec:zf}

To reduce the computational complexity of the SOCP-based design, we further develop a ZF-based approach that admits explicit peak-power expressions. Define the channel matrix
\begin{equation}
\mathbf H(\mathbf t)
\triangleq
\begin{bmatrix}
a&b\\c&d
\end{bmatrix}
=
\begin{bmatrix}
h_{1,1}(t_1)&h_{2,1}(t_2)\\
h_{1,2}(t_1)&h_{2,2}(t_2)
\end{bmatrix},
\nonumber
\end{equation}
where $\Delta\triangleq ad-bc$ and
$\zeta\triangleq\sqrt{\gamma\sigma^2}$.
We consider $\Delta\neq0$ in the following ZF designs.

\subsubsection{One-Sided Cooperation}

We first consider $w_{1,2}=0$, where BS 2 additionally
transmits user 1's data.
By imposing $cw_{1,1}+dw_{2,1}=0$, the interference
caused by user 1's stream at user 2 is eliminated.
The resulting partial-ZF precoder is
\begin{align*}
\mathbf W_{\rm L}
=
\begin{bmatrix}
du/\Delta&0\\
-cu/\Delta&\zeta/d
\end{bmatrix},
\end{align*}
where $u=\zeta\sqrt{1+\gamma|b/d|^2}$.
It follows that
\begin{align*}
\mathbf H\mathbf W_{\rm L}
=
\begin{bmatrix}
u&b\zeta/d\\
0&\zeta
\end{bmatrix}.
\end{align*}
Thus, both users achieve the target SINR $\gamma$,
and the required peak BS power can be written as follows:
\begin{align*}
\Phi_{\rm L}(\mathbf t)
=
\frac{\gamma\sigma^2}{|\Delta|^2}
\max\left\{
A_{\rm L},
\frac{|c|^2A_{\rm L}+|\Delta|^2}{|d|^2}
\right\},
\end{align*}
where $A_{\rm L}\triangleq|d|^2+\gamma|b|^2$.

Similarly, for the opposite cooperation direction
$w_{2,1}=0$, the partial-ZF precoder is
\begin{align*}
\mathbf W_{\rm U}
=
\begin{bmatrix}
\zeta/a&-bv/\Delta\\
0&av/\Delta
\end{bmatrix},
\end{align*}
where $v=\zeta\sqrt{1+\gamma|c/a|^2}$.
The corresponding required peak BS power can be written as follows:
\begin{align*}
\Phi_{\rm U}(\mathbf t)
=
\frac{\gamma\sigma^2}{|\Delta|^2}
\max\left\{
\frac{|b|^2A_{\rm U}+|\Delta|^2}{|a|^2},
A_{\rm U}
\right\},
\end{align*}
where $A_{\rm U}\triangleq|a|^2+\gamma|c|^2$.
Both partial-ZF constructions achieve the target
SINRs with equality.

The minimum required peak power under one-sided
partial-ZF transmission is therefore
\begin{align*}
Q_{1,\rm pZF}^\star
=
\min\nolimits_{t_1,t_2\in\mathcal T}
\min\{\Phi_{\rm L}(\mathbf t),
\Phi_{\rm U}(\mathbf t)\}.
\end{align*}
We optimize the two cooperation directions
separately and select the better solution.

\subsubsection{Full Cooperation}

Under full cooperation, conventional ZF eliminates
the inter-user interference at both users.
The corresponding precoding matrix is given by
\begin{align*}
\mathbf W_{\rm F}
=
\zeta\mathbf H^{-1}
=
\frac{\zeta}{\Delta}
\begin{bmatrix}
d&-b\\
-c&a
\end{bmatrix}.
\end{align*}
Since $\mathbf H\mathbf W_{\rm F}=\zeta\mathbf I_2$,
both users achieve the target SINR $\gamma$.
The required peak BS power is given by
\begin{align*}
\Phi_{\rm F}(\mathbf t)
=
\frac{\gamma\sigma^2}{|\Delta|^2}
\max\left\{
|d|^2+|b|^2,
|c|^2+|a|^2
\right\}.
\end{align*}
The corresponding PA-placement problem is
\begin{align*}
Q_{2,\rm ZF}^\star
=
\min\nolimits_{t_1,t_2\in\mathcal T}
\Phi_{\rm F}(\mathbf t).
\end{align*}

\subsubsection{PA Placement Optimization}

The above designs yield three explicit peak-power
objectives:
$\Phi_{\rm L}(\mathbf t)$,
$\Phi_{\rm U}(\mathbf t)$, and
$\Phi_{\rm F}(\mathbf t)$.
Each objective depends only on the two PA positions.
We therefore apply the element-wise AO method
developed in Section~\ref{subsec:no_cooperation}.
Specifically, for
$\nu\in\{{\rm L},{\rm U},{\rm F}\}$,
the PA positions are updated sequentially as
\begin{align*}
t_k\leftarrow
\argmin\nolimits_{t\in\mathcal Q\cup\{t_k\}}
\Phi_\nu(\mathbf t|_{t_k=t}).
\end{align*}
Candidates with $\Delta=0$ are excluded.
Since the current PA position is included in each
search, the objective value is non-increasing
over successive updates.
\vspace{-5pt}
\begin{remark}
Unlike the SOCP-based design, the proposed
ZF-based approach only requires evaluations of
closed-form power expressions during PA placement.
However, the additional interference-nulling
constraints may increase the required transmit power.
Therefore, the ZF-based solutions provide
low-complexity feasible designs rather than
globally optimal solutions to the original problem.
\end{remark}
\vspace{-5pt}
\subsection{Computational Complexity Analysis}
\label{subsec:complexity}

Let $I_0$, $I_{\rm A}$, and $I_{\rm Z}$ denote the numbers of AO iterations for independent transmission, SOCP-based cooperation, and ZF-based cooperation, respectively. Their computational complexities are $\mathcal O(I_0Q)$, $\mathcal O(I_{\rm A}(C_{\rm SOCP}+I_{\rm P}Q))$, and $\mathcal O(I_{\rm Z}Q)$, respectively, where $I_{\rm P}$ is the number of inner PA-position AO iterations and $C_{\rm SOCP}=\mathcal O(n^3\log(1/\epsilon))$ is an interior-point complexity estimate for one SOCP, with $n=7$ and $n=9$ real optimization variables for
one-sided and full cooperation, respectively, and $\epsilon$ the solution accuracy. Thus, both cooperative designs scale linearly with $Q$, while the ZF-based design avoids the additional SOCP complexity.

\section{Numerical Results}
\label{sec:numerical}

In this section, we evaluate the cooperation cost of PASS and compare it with the conventional fixed-antenna system, where each antenna is deployed at the cell center. Unless otherwise stated, the carrier frequency is $f_{\rm c}=28$ GHz, the noise power is $\sigma^2=-90$ dBm, the effective refractive index is $n_{\rm{eff}}=1.4$, and the waveguide height is $H=3$ m. Two adjacent cells are considered, each with dimensions $D_x=80$ m and $D_y=20$ m and one uniformly distributed user, as illustrated in {\figurename}~\ref{Figure: System_Model}. The proposed SOCP-based design is initialized using the ZF-based solutions. Each one-dimensional PA search employs a uniform grid of $Q=2001$ points. The target rate is $R_{\rm th}=\log_2(1+\gamma)$, where Gaussian signaling is assumed and interference is treated as noise. The power budget $P_{\max}$ is set to the $60$th percentile of the minimum peak transmit power required by the conventional fixed-antenna system with ZF-based full cooperation. This value is estimated using $10^5$ Monte-Carlo realizations and recalculated as $D_x$ or $R_{\rm th}$ varies. The cooperation cost is capped at $C^\star=2$ for all realizations where the SINR requirements cannot be satisfied.

\begin{figure}[!t]
\centering
    \subfigure[$R_{\rm{th}}=3$ bps/Hz.]
    {
        \includegraphics[width=0.44\textwidth]{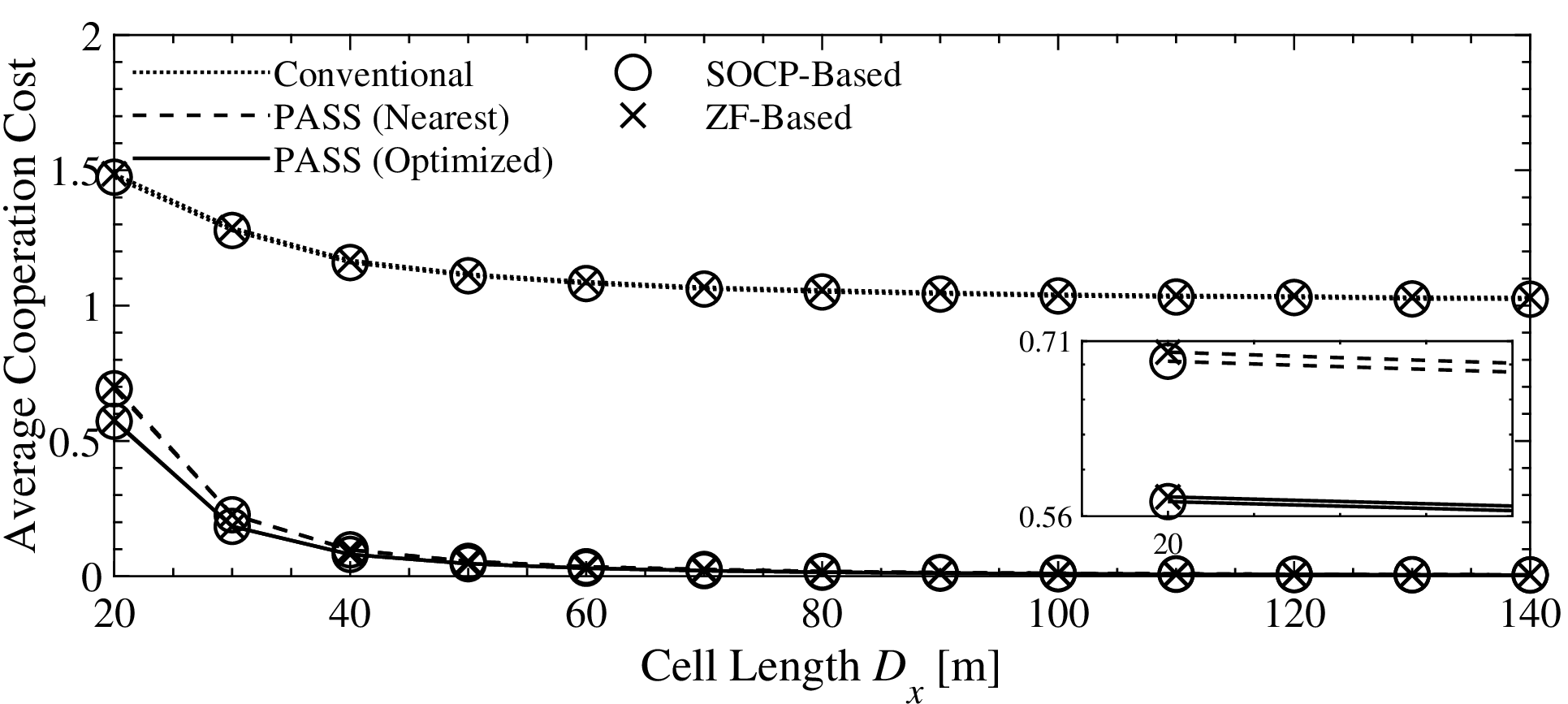}
	   \label{fig:Figure1a}
    }
    \subfigure[$D_x=80$ m.]
    {
        \includegraphics[width=0.44\textwidth]{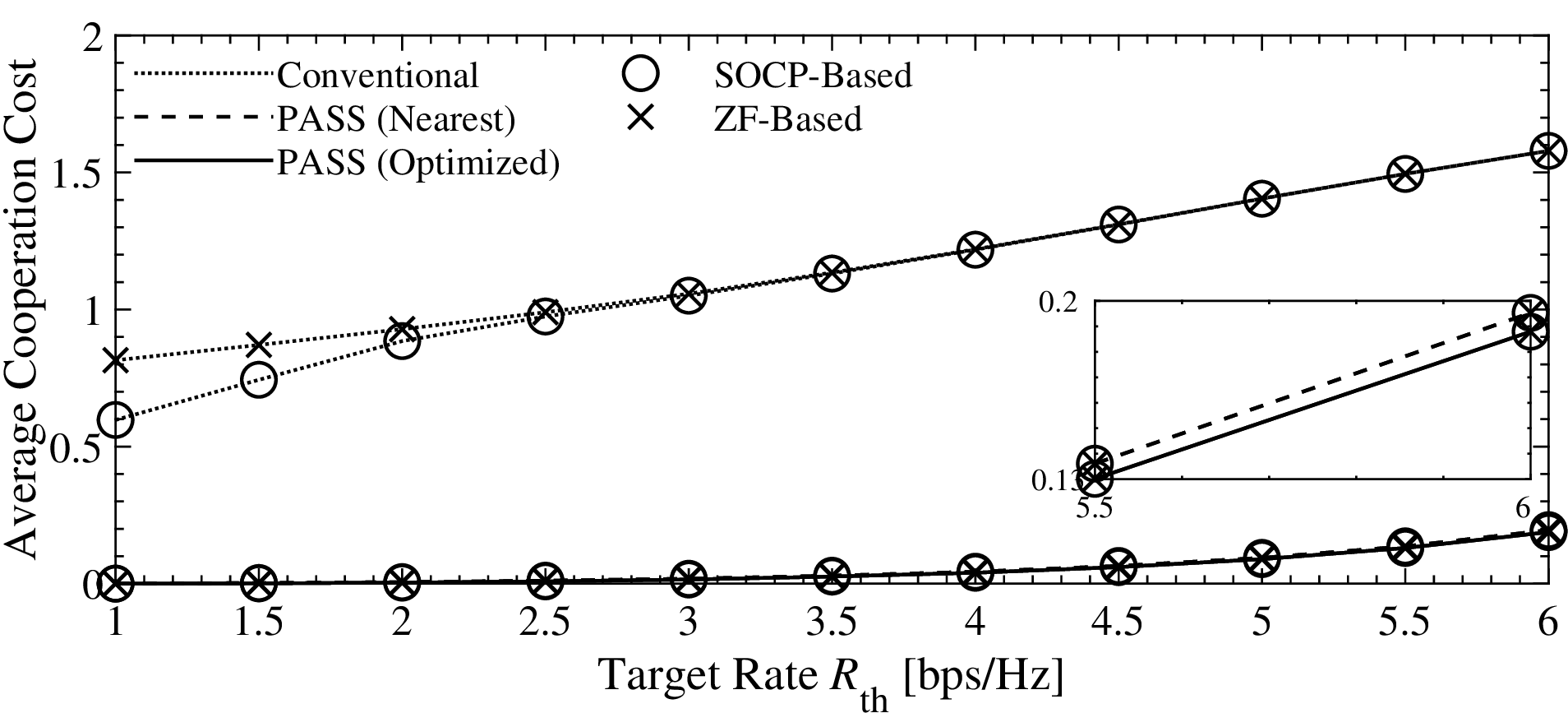}
	   \label{fig:Figure1b}
    }
\caption{Average cooperation cost.}
\label{fig:Figure1}
\vspace{-15pt}
\end{figure}

{\figurename}~\ref{fig:Figure1a} compares the average cooperation cost of the conventional fixed-antenna system, nearest-PA placement, and optimized PA placement under SOCP-based and ZF-based precoding. Here, nearest-PA placement positions each PA at the projection of its intended user onto the corresponding waveguide, i.e., $t_k=x_k^{\rm U}-x_k^{\rm C}$ for $k\in\{1,2\}$. All four PASS schemes achieve substantially lower cooperation costs than their conventional counterparts. This improvement is primarily attributed to the spatial flexibility of PASS, which enhances the desired channels and alleviates ICI. The proposed optimized PA placement further outperforms nearest-PA placement, particularly for small cell lengths. For example, at $D_x=20$ m, the average cooperation cost is reduced from $0.69$ to $0.57$ under SOCP-based precoding. As $D_x$ increases, the cooperation costs decrease for all schemes, while the performance gap between the two PA placement strategies becomes negligible. Furthermore, ZF-based precoding achieves performance close to that of SOCP-based precoding, which demonstrates its effectiveness for cooperation reduction. {\figurename}~\ref{fig:Figure1b} shows the average cooperation cost versus the target rate $R_{\rm th}$. As expected, the cooperation cost increases with $R_{\rm th}$, since higher SINR requirements make independent transmission more difficult. Nevertheless, PASS consistently outperforms the conventional fixed-antenna system. Remarkably, even at $R_{\rm th}=6$ bps/Hz, the proposed PASS achieves an average cooperation cost below $0.2$, compared with approximately $1.6$ for the conventional fixed-antenna system. The relatively small difference between nearest-PA and optimized PA placement further indicates that simple geometry-based PA placement can capture most of the cooperation reduction benefits. This highlights the practicality of PASS, as substantial cooperation savings can be achieved without sophisticated PA placement optimization.

\begin{figure}[!t]
\centering
    \subfigure[$R_{\rm{th}}=3$ bps/Hz.]
    {
        \includegraphics[height=0.17\textwidth]{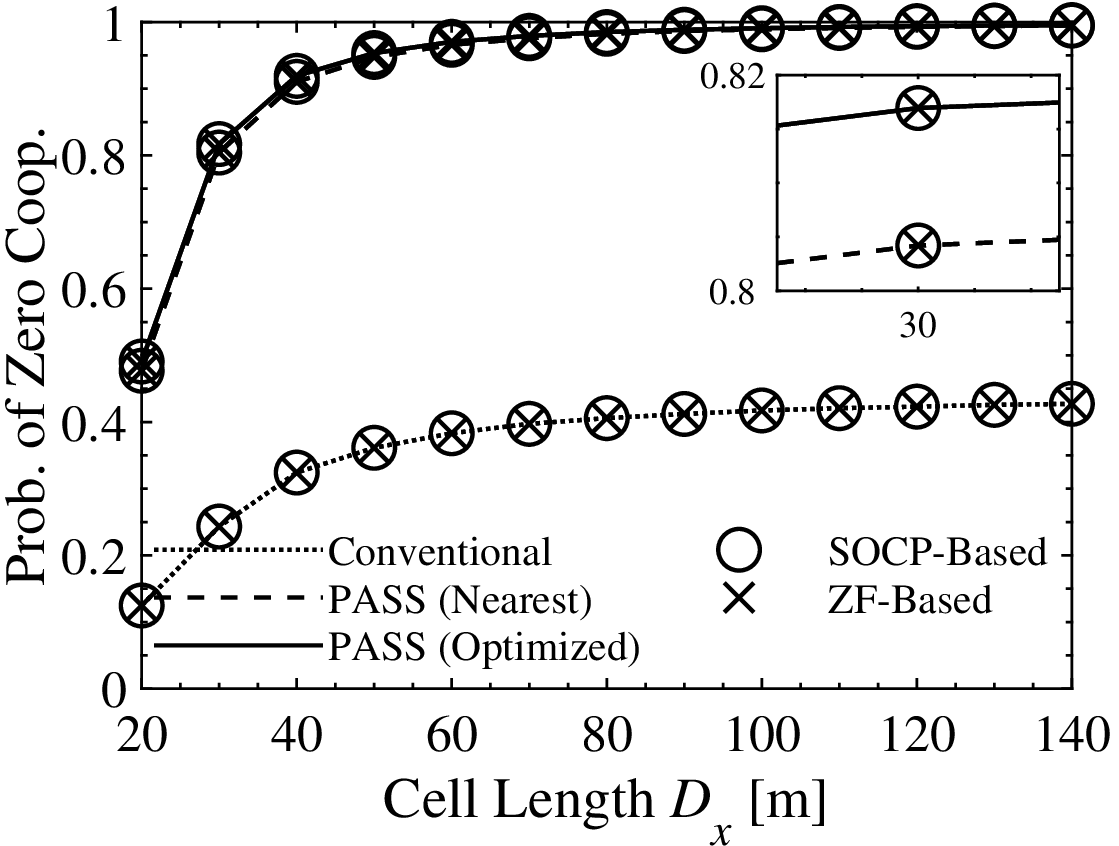}
	   \label{fig:Figure2a}
    }
    \subfigure[$D_x=80$ m.]
    {
        \includegraphics[height=0.17\textwidth]{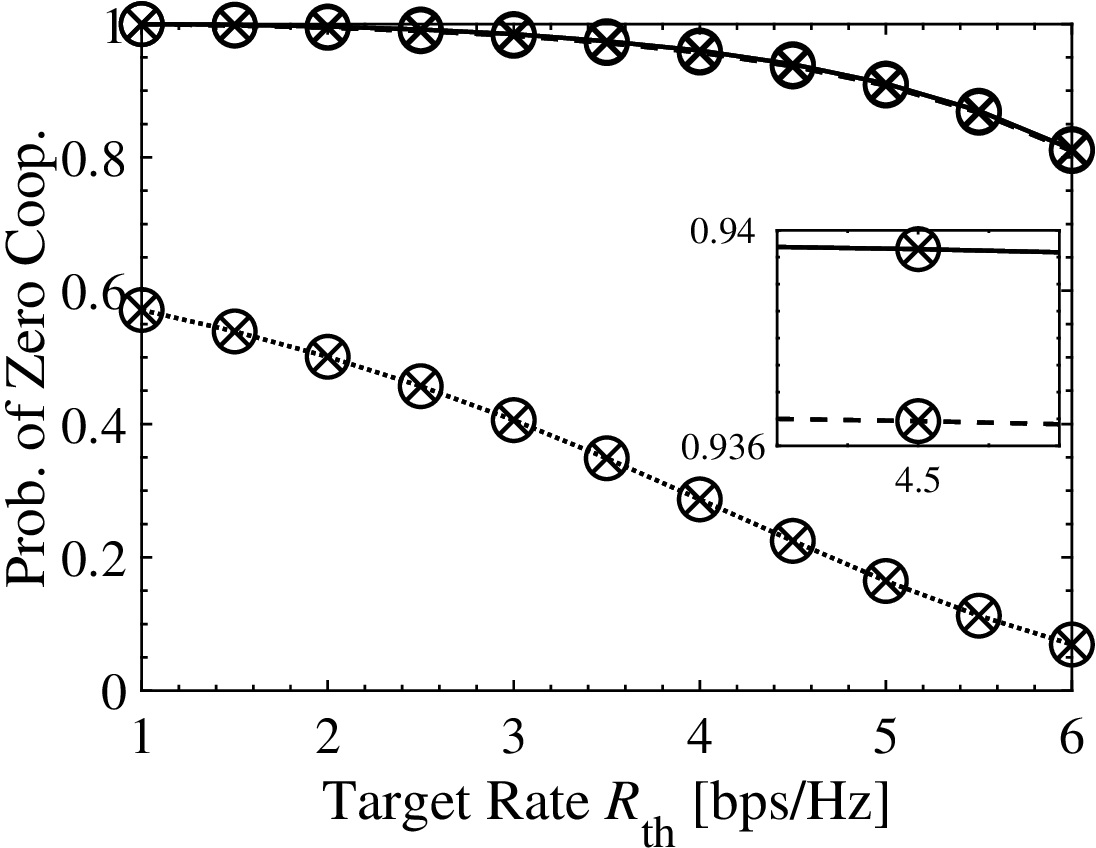}
	   \label{fig:Figure2b}
    }
\caption{Probability of zero cooperation $\Pr(C=0)$.}
\label{fig:Figure2}
\vspace{-15pt}
\end{figure}

To further characterize the potential of PASS for independent transmission, {\figurename}~\ref{fig:Figure2} presents the probability of zero cooperation, $\Pr(C=0)$. As shown in {\figurename}~\ref{fig:Figure2a}, this probability increases with $D_x$ for all schemes, but PASS achieves a substantially higher probability than the conventional fixed-antenna system. {\figurename}~\ref{fig:Figure2b} shows that $\Pr(C=0)$ decreases with $R_{\rm th}$ due to the increasingly stringent SINR requirements. Nevertheless, the proposed PASS maintains a probability above $0.8$ even at $R_{\rm th}=6$ bps/Hz, whereas the conventional fixed-antenna system achieves less than $0.1$. These results demonstrate that PASS can substantially reduce inter-BS data sharing and enable a transition from cooperative transmission to independent cell operation.

\section{Conclusion}
\label{sec:conclusion}

This article has demonstrated that flexible PA placement can substantially reduce multi-cell cooperation requirements and transform cooperative ``choirs'' into independent ``solos''. Moreover, nearest-PA placement achieves performance close to that of the proposed joint design, which makes it an attractive low-complexity alternative.

\bibliographystyle{IEEEtran}
\bibliography{mybib}
\end{document}